\documentclass[fleqn,twoside,twocolumn,nofootinbib,showkeys]{revtex4}
\begin{document}

\title{Comment on  perihelion advance due to cosmological constant}

\author{S.~S.~Ovcherenko} 
\affiliation{Novosibirsk State University, 630 090, Novosibirsk, Russia}

\author{Z.~K.~Silagadze}
\affiliation{Budker Institute of Nuclear Physics and
Novosibirsk State University, 630 090, Novosibirsk, Russia} 
\email{silagadze@inp.nsk.su}           

\begin{abstract}
We comment on the recent paper ``Note on the perihelion/periastron 
advance due to cosmological constant'' by H.~Arakida (Int. J. Theor. Phys. 
52 (2013) 1408-1414) and provide simple derivations both of the main result
of this paper and of the Adkins-McDonnell's precession formula, on which
this  main result is based.
\end{abstract}
\maketitle

\section{Introduction}
Recently Hideyoshi Arakida in the interesting paper \cite{1} clarified some 
confusion existing in the literature concerning the eccentricity dependence
of  the perihelion/periastron advance of celestial bodies due to the 
cosmological constant $\Lambda$. He showed that the correct expression
for the perihelion/periastron shift per period is
\begin{equation}
\Delta \Theta_p=\frac{\pi c^2\Lambda a^3}{GM}\sqrt{1-e^2},
\label{eq1}
\end{equation}
where $a$ is the semi-major axis of the orbit and $e$ is the eccentricity. 
This result was obtained in \cite{1} by the help of the general formula
\begin{equation}
\Delta \Theta_p=-\frac{2p}{\alpha e^2}\int\limits_{-1}^1\frac{dV(z)}{dz}\,
\frac{z}{\sqrt{1-z^2}}\,dz
\label{eq2}
\end{equation}
for the perihelion/periastron shift per period due to a small central-force 
perturbation $$V(z)=V(r(z))=V\left(\frac{p}{1+ez}\right),\;\;
\;p=a(1-e^2),$$ to the Newtonian 
potential $V_0(r)=-\alpha/r$, $\alpha=GMm$. The formula (\ref{eq2}) was first 
obtained in \cite{2}. Note that, contrary to \cite{1}, but in accord with 
\cite{2}, our $V(r)$ is the perturbation potential energy, not the 
perturbation potential as in \cite{1}, and therefore it  includes the mass of 
the orbiting particle $m$. 
 
Now we demonstrate that suitably modified Landau and Lifshitz's approach 
\cite{3} allows to simply derive both (\ref{eq1}) and (\ref{eq2}).

\section{Landau and Lifshitz's approach}
Landau and Lifshitz provide \cite{3} the following expression for 
$\Delta \Theta_p$ (see solution of the Problem 3 in $\S 15$):
\begin{equation}
\Delta \Theta_p=\frac{\partial}{\partial L}\int\limits_{r_{min}}^{r_{max}}
\frac{2mV(r)}{\sqrt{2m\left(E+\frac{\alpha}{r}\right)-\frac{L^2}{r^2}}}\,dr,
\label{eq3}
\end{equation}
where $L$ is the angular momentum, $E$ is the total energy, and the 
integration is over the unperturbed Keplerian orbit. A simple derivation of 
(\ref{eq3}) can be found in \cite{3}. 
Using
$$\frac{dr}{\sqrt{2m\left(E+\frac{\alpha}{r}\right)-\frac{L^2}{r^2}}}=
\frac{1}{m}\,dt,$$
and extending the integration over the whole orbital period $T$, it is 
convenient to rewrite (\ref{eq3}) in the following form \cite{4}
\begin{equation}
\Delta \Theta_p=\frac{\partial}{\partial L}\int\limits_0^T V(r(t))\,dt=
\frac{\partial}{\partial L}\left(T<V>\right),
\label{eq4}
\end{equation}
where
$$<V>=\frac{1}{T}\int\limits_0^T V(r(t))\,dt$$
is the time-average value of the perturbation potential energy over the 
unperturbed orbit. Note that this time-averaged value is a function of $L$ 
and $E$ and it is the total energy $E$ that is to be kept 
constant when taking the partial derivative $\partial/\partial L$ in 
(\ref{eq4}).

To apply (\ref{eq4}) to the problem considered in \cite{1}, with the 
perturbation potential energy
\begin{equation}
V(r)=-\frac{1}{6}\Lambda mc^2r^2,
\label{eq5}
\end{equation}
let us use the following parametrization of the unperturbed motion on the 
Keplerian ellipse \cite{3}:
\begin{equation}
t=\sqrt{\frac{ma^3}{\alpha}}(\xi-e\,\sin{\xi}),\;\;\;
r=a(1-e\,\cos{\xi}),
\label{eq6}
\end{equation}
where the parameter $\xi$ changes from $0$ to $2\pi$. As a result, we get
\begin{equation}
\Delta \Theta_p=-\frac{1}{6}\Lambda mc^2a^2\sqrt{\frac{ma^3}{\alpha}}
\frac{\partial}{\partial L}\int\limits_0^{2\pi}(1-e\,\cos{\xi})^3\,d\xi,
\label{eq7}
\end{equation}
and after the elementary evaluation of the integral,
\begin{equation}
\Delta \Theta_p=-\frac{\pi}{3}\Lambda mc^2a^2\sqrt{\frac{ma^3}{\alpha}}
\frac{\partial}{\partial L}\left(1+\frac{3}{2}e^2\right).
\label{eq8}
\end{equation}
But
$$e^2=1+\frac{2EL^2}{m\alpha^2}=1-\frac{L^2}{m\alpha a},$$
and, therefore,
\begin{equation}
\frac{\partial e^2}{\partial L}=-\frac{2L}{m\alpha a}=-2\sqrt{\frac{1-e^2}
{m\alpha a}},
\label{eq9}
\end{equation}
which together with (\ref{eq8}) imply the validity of (\ref{eq1}):
\begin{equation}
\Delta \Theta_p=\frac{\pi\Lambda mc^2a^3}{\alpha}\,\sqrt{1-e^2}=
\frac{\pi\Lambda c^2a^3}{GM}\,\sqrt{1-e^2}.
\label{eq10}
\end{equation}

It remains to clarify how the Adkins-McDonnell's precession formula 
(\ref{eq2}) can be obtained from (\ref{eq4}). Using again the parametrization
(\ref{eq6}), we can write (with $r(\xi)=a(1-e\,\cos{\xi})$)
\begin{equation}
\Delta \Theta_p=\sqrt{\frac{ma^3}{\alpha}}\frac{\partial}{\partial L}
\int\limits_0^{2\pi}V(r(\xi))\,(1-e\,\cos{\xi})\,d\xi.
\label{eq11}
\end{equation}
Because $\cos{(2\pi-\xi)}=\cos{\xi}$, this can be rewritten in the form
\begin{equation}
\Delta \Theta_p=2\sqrt{\frac{ma^3}{\alpha}}\frac{\partial}{\partial L}
\int\limits_0^{\pi}V(r(\xi))\,(1-e\,\cos{\xi})\,d\xi.
\label{eq12}
\end{equation}
Now let us apply the Leibniz integral rule (differentiation under the integral 
sign) to get
\begin{eqnarray}&&
\Delta \Theta_p=2\sqrt{\frac{ma^3}{\alpha}}\,\frac{\partial e}{\partial L}
\times \nonumber \\ &&
\left [\int\limits_0^{\pi}V^\prime(r(\xi))\,(-a\cos{\xi})
(1-e\,\cos{\xi})\,d\xi+I\right],
\label{eq13}
\end{eqnarray}
where
\begin{equation}
I=-\int\limits_0^{\pi}V(r(\xi))\,\cos{\xi}\,d\xi=
-\int\limits_0^{\pi}V(r(\xi))\,d(\sin{\xi}),
\label{eq14}
\end{equation}
and $V^\prime(r)=\frac{dV(r)}{dr}$.
Integration by parts, along with
$$\frac{dV(r(\xi))}{d\xi}=ea\,V^\prime(r(\xi))\,\sin{\xi},$$
allows to rewrite (\ref{eq14}) in the form
\begin{equation}
I=ea\,\int\limits_0^{\pi}V^\prime(r(\xi))\,\sin^2{\xi}\,d\xi.
\label{eq15}
\end{equation}
Substituting (\ref{eq15}) into (\ref{eq13}), we get
\begin{equation}
\Delta \Theta_p=2\sqrt{\frac{ma^3}{\alpha}}\,\frac{\partial e}{\partial L}\,
a\,\int\limits_0^{\pi}(e-\cos{\xi})\,V^\prime(r(\xi))\,d\xi.
\label{eq16}
\end{equation}
At this stage, let us introduce a new integration variable $z$:
\begin{equation}
z=\frac{\cos{\xi}-e}{1-e\,\cos{\xi}},\;\;\;\cos{\xi}=\frac{z+e}{1+e\,z}.
\label{eq17}
\end{equation}
It follows from (\ref{eq17}) that
\begin{eqnarray} &&
e-\cos{\xi}=-\frac{(1-e^2)z}{1+e\,z},\;\;
1-e\,\cos{\xi}=\frac{1-e^2}{1+e\,z}, \nonumber \\ &&
\sin^2{\xi}=\frac{(1-z^2)(1-e^2)}
{(1+e\,z)^2},
\label{eq18}
\end{eqnarray}
and
\begin{equation}
d\xi=-\frac{1-e^2}{(1+e\,z)^2}\,\frac{1}{\sin{\xi}}\,dz=
-\frac{\sqrt{1-e^2}}{(1+e\,z)\sqrt{1-z^2}}\,dz.
\label{eq19}
\end{equation}
Besides $\frac{dV(r(z))}{dz}$ equals to
\begin{equation}
\frac{d}{dz}V\left(\frac{a(1-e^2)}{1+e\,z}\right)=
V^\prime(r(z))\left(-\frac{a\,e\,(1-e^2)}{(1+e\,z)^2}\right),
\label{eq20}
\end{equation}
that allows to express $V^\prime(r(\xi))=V^\prime(a(1-e\,\cos{\xi}))=
V^\prime(r(z))$ in terms of $\frac{dV(r(z))}{dz}$. After taking all these 
relations into account, (\ref{eq16}) becomes
\begin{equation}
\Delta \Theta_p=2\sqrt{\frac{ma^3}{\alpha}}\,\frac{1}{e}
\frac{\partial e}{\partial L}\,\sqrt{1-e^2}\int\limits_{-1}^1\frac{dV(r(z))}
{dz}\frac{z\,dz}{\sqrt{1-z^2}},
\label{eq21}
\end{equation}
and this coincides to the Adkins-McDonnell's precession formula (\ref{eq2}),
because, due to  (\ref{eq9}),
$$2\sqrt{\frac{ma^3}{\alpha}}\,\frac{1}{e}\,\frac{\partial e}{\partial L}\,
\sqrt{1-e^2}=-2\frac{a(1-e^2)}{\alpha\,e^2}=-\frac{2p}{\alpha\,e^2}.$$

\section{Concluding remarks}
Various approaches to account the influence of the cosmological constant
on the celestial dynamics can be found in  references cited in \cite{1}.
Kotkin and Serbo's variant (\ref{eq4}) of the Landau and Lifshitz's precession 
formula and the parametrization (\ref{eq6}) of the unperturbed 
motion on the Keplerian ellipse provide, probably, the simplest way to
calculate   the perihelion/periastron advance of celestial bodies due to the 
cosmological constant in the framework of the Schwarzschild-de Sitter 
(Kottler) space-time. This approach also allows a simple derivation of the 
Adkins-McDonnell's precession formula (\ref{eq2}) (another simple derivation 
of this formula, based on the precession of the Hamilton's vector, was 
given in \cite{5}).

\acknowledgments
The work is supported by the Ministry of Education and Science of the Russian 
Federation and in part by Russian Federation President Grant for the support 
of scientific schools NSh-2479.2014.2.

\end{document}